# Volumetric parametrization from a level set boundary representation with PHT-splines

Chiu Ling Chan[c], Cosmin Anitescu[c], Timon Rabczuk[a,b,c,*]

[a]*Division of Computational Mechanics, Ton Duc Thang University, Ho Chi Minh City, Vietnam.*
[b]*Faculty of Civil Engineering, Ton Duc Thang University, Ho Chi Minh City, Vietnam*
[c]*Institute of Structural Mechanics, Bauhaus Universität Weimar, Germany*

**Abstract**

A challenge in isogeometric analysis is constructing analysis-suitable volumetric meshes which can accurately represent the geometry of a given physical domain. In this paper, we propose a method to derive a spline-based representation of a domain of interest from voxel-based data. We show an efficient way to obtain a boundary representation of the domain by a level-set function. Then, we use the geometric information from the boundary (the normal vectors and curvature) to construct a matching $C^1$ representation with hierarchical cubic splines. The approximation is done by a single template and linear transformations (scaling, translations and rotations) without the need for solving an optimization problem. We illustrate our method with several examples in two and three dimensions, and show good performance on some standard benchmark test problems.

## 1. Introduction

The premise of isogeometric analysis (IGA) has been to provide a numerical method by which the basis functions used in CAD to represent the geometry are also used to compute the approximate solution fields [12]. Some potential obstacles in this direction are that the CAD geometry is usually provided in boundary form and that it often uses Boolean operations (trim curves and surfaces), which makes it unsuitable for direct use in analysis. Moreover, there is demand for generating analysis-suitable models from sources other than CAD [38], such as voxelized data provided by CT-scans or other specialized imaging equipment. The latter in particular has important uses in medical fields (for example for the study of the mechanical characteristics of implants) or applications related to digital image correlation.

A challenge in volumetric parametrization is generating a mesh for complex shapes when little *a priori* structure is known about the domain. For example, a tensor-product basis can represent easily "boxy" domains leading to accurate geometry descriptions with few degrees of freedom. However, a tensor-product basis is less suitable for domains with inclusions or other irregular features.

---

[*]Corresponding author
  *Email address:* `timon.rabczuk @ tdt.edu.vn` (Timon Rabczuk)

B-Splines and NURBS, which are often used in CAD systems, are the most common basis in IGA, provided that a geometry description of the interior of the domain is given. If only a boundary representation is provided, but the domain can be approximated by a (curved) rectangle or cube, then a volumetric description can be constructed using a Gordon-Coons parametrization [13]. However, in general, it cannot be guaranteed that the obtained geometry will be suitable for analysis. For more complex domains, typically a multi-patch description is required for which the connectivity information between the patches needs to be managed. There is usually only $C^0$ continuity at the patch boundaries, and higher order continuity is more cumbersome to enforce. Another drawback is that due to the tensor product structure, refinements propagate throughout the patch or even the domain, so true local refinement is difficult to realize. Thus, there are many research papers discussing various improvements or modifications aimed at constructing complex solid meshes that are suitable for analysis. For example, an adaptive approach is presented in [28], and a local refinement method is given in [15]. In [19], trivariate NURBS representations of composite panels are discussed. A discussion on using B-splines focused on obtaining meshes of complex geometries through an optimization method is presented in [34]. Another approach us using an Isogeometric Boundary Element Method [30], which only requires a discretization of the boundary to represent the geometry and conduct the analysis. More recent advances regarding multi-patch parametrizations are given in [5] and [2].

Other splines which are widely used in IGA and provide more flexibility are T-splines. They were introduced in [29] to allow meshes with T-junctions, thus, they have better local refinement properties compared to B-splines or NURBS. Bazilevs et al. [4] discussed the application of T-splines in IGA and a parametrization using volumetric T-splines is presented in [16]. In [33], a method for converting a quadrilateral mesh into a T-spline surface is introduced. It is highlighted in [8] that local refinement with T-splines can produce more accurate results, and the use of adaptive $h$-refinement with T-splines is discussed.

Polynomial Splines over Hierarchical T-meshes (PHT Splines) [7] share some similarities with hierarchical B-splines and T-splines, and have more flexible local refinement properties. Thus, they are quite suitable for use in the mesh generation for complex shapes. Isogeometric analysis methods using PHT-splines and various refinement strategies have been introduced in [20, 22, 32, 21]. In this work, we propose an automatic method to construct a PHT-spline model that matches the geometry features of a given physical domain.

While the proposed method is applicable to other types of input data, such as a boundary spline representations or trimmed NURBS surfaces, we focus on obtaining volumetric descriptions of objects of interest from image data. We present a fast and efficient way to convert the voxelized volume description to an implicit level-set description of the boundary. Then the geometric information extracted from the level-set function (in particular the normal to the boundary and the curvature) is used to construct an explicit spline parametrization of the volume. We use non-uniform refinements to ensure a good match of the parametrized domain with the given geometry, while using up to several orders of magnitude fewer elements compared to the number of voxels in the input data.

The approach used in this work is in some ways related to other methods for dealing with complex boundaries, such as the Immersed Boundary Method (IBM), Immerse Finite Element Method (IFEM) [37], and the Finite cell method (FCM) [24]. IBM and IFEM are widely used in fluid mechanics. However, they tend to introduce geometric modeling errors which negatively impacts the accuracy of the solutions. The idea of FCM was to apply adaptive analysis over a mesh of simple geometry, using a highly refined integration mesh to capture the limits of the domain. Nevertheless, this approach can lead to ill-conditioning if the elements cut by the boundary are too small, and the accuracy of the method is typically limited by the approximate geometry representation. Another related method is the Immersed Particle Method [26], which is based on the same concept as IFEM but has been developed for modelling fluid-structure interaction. In [36], a surface reconstruction from scanned image using B-spline is presented. The idea of this method is to project the grid points extracted from voxel data in the normal direction onto the surface. The Weighted Extended B-Splines (WEB-splines) [14] provide another interesting approach for representing free form geometries in the context of finite element simulations. In this method, B-splines are marked as inner or outer based on the location of their support and are weighted by a distance function. The resulting basis is non-polynomial and not fully compatible with existing CAD implementation. In [27], an immersed B-spline method was proposed to interpolate the domain boundaries using an isoparametric basis. However, the normal and curvature information of the surface is not considered and seems more difficult to incorporate along with non-uniform refinement due to the $C^2$ continuity of the basis.

Also related to our work, we can mention Cartesian grid methods, which is another approach for determining solid meshes. These methods are also suitable for analyzing voxel-based data, and the idea is to use the voxel structure as the computational domain. However, because many 3D images obtained using current technology can contain hundreds of millions of voxels, these methods are typically restricted to linear finite element approximations. Meshfree particle methods which have been developed based on Cartesian grids are introduced in [25]. In the recent method presented in [18], transfinite mapping functions are used at the elements cut by the boundary to avoid integration error. Reparameterization methods [35] are another approach to volumetric parametrization. Boundary reparameterization is performed as a pre-processing step to improve the isoparametric structure. The optimal control points and weights of the reparameterized surface are then obtained using a variational harmonic metric. The idea of using harmonic mappings to determine a planar parametrization is also explored in [9].

An appealing advantage of our method is that a complex domain can be meshed automatically by using only one patch of PHT-splines. In addition, an initial boundary representation such as a boundary triangulation is not necessary. Thus the method is in this sense self-contained. The key aspect of our proposed method is that we match the boundary of the solid in the physical domain by approximating the boundary with a single template consisting of a circular-arc in 2D and a sphere-surface in 3D. This approach allows the construction of $C^1$ representation of a domain of arbitrary complexity using only translations, rotations, and scaling operations at the boundary nodes. The resulting geometric mapping may contain non-regular points, which can also be found on a circle or sphere parametrization based on a tensor-product basis. However, different from the immersed boundary or weighted B-spline method, the domain representation is based on just Bernstein-Bézier polynomials with the usual control points and defined on a hierarchical mesh. Moreover, the reduced overlap between the basis functions allows for more local and granular control over the parametrization at the mesh vertices.

The paper is structured as follows: a brief introduction of the level set function and the computation of geometry features from the physical domain are discussed in Section 2. In Section 3, we introduce the PHT-splines and some computational details regarding the 2D and 3D PHT-splines. The volumetric parametrization algorithm is discussed in Section 4, where we include the details of using the circle and spherical templates to modify the boundary vertices. The numerical results are presented in Section 5, and some concluding remarks are given in the last section.

## 2. B-spline level set

In this section, we describe how to efficiently employ a level set method [23] to represent the boundary of an image. Suppose a voxelized image is represented by a discontinuous gray scale function $g : \Omega_I \to \mathbb{R}$ as follows:

$$g(X) = \begin{cases} I_1 & : X \in \Omega_I^1 \\ I_2 & : X \in \Omega_I^2 \\ \vdots \\ I_M & : X \in \Omega_I^M \end{cases}$$

where $\Omega_I \subset \mathbb{R}^d$ is the image domain, $X$ represents the voxel position and $I_j$, $j = 1, 2, \ldots, M$ is the gray scale value in the image of each voxel $\{\Omega_I^e\}_{e=1}^M$. We also assume that the domain of interest (*computational domain*) in the image has a lower gray scale intensity than the rest of the image. Then it can be approximated as the set $\{X : g(X) \leq T\}$, where $T$ is a user-definable threshold which determines the boundary of the domain. Moreover, if $\hat{g}(X)$ is a suitably smoothed representation of $g(X)$, then the boundary is represented by the contour line $\{X : \hat{g}(X) = T\}$ and the following fundamental properties of level set function hold (in both 2D and 3D):

1. The unit normal vector to the contour line (boundary) at point $X$ is given by $\frac{\nabla \hat{g}}{|\nabla \hat{g}|}$.

2. The curvature of the contour line is $\nabla \cdot \frac{\nabla \hat{g}}{|\nabla \hat{g}|}$.

We note that while it is possible to obtain a contour line directly from pixel data, the boundary may contain jagged edges or noise artifacts from the image and the approximate gradients computed from the pixel intensity values may be inaccurate.

In order to obtain a smooth representation of the boundary, we compute an approximation of $g(X)$ using a B-spline level set function:

$$f(X) = \sum_{j=1}^{M} N_{j,p}(X) c_j, \quad j = 1, \ldots, M, \tag{1}$$

where $N_{j,p}(X)$ and $c_j$ represent the basis functions and scalar coefficients respectively. These functions are defined by uniform knot vectors with non-repeated integer knot values. In [31], a convolution strategy for computing the coefficients $c_j$ which gives a smooth representation of the gray scale function $g(X)$ has been proposed. The formula used is:

$$c_j = \frac{\int_{\Omega_I} N_{j,p}(X)\, g(X)\, dX}{\int_{\Omega_I} N_{j,p}(X)\, dX}, \quad j = 1, ..., M. \tag{2}$$

We note that for $d > 1$, $N_{j,p}(X)$ is a tensor product of 1D B-Splines which are translation invariant. For example, when $d = 2$ we can write: $N_{j,p}(X) = N_p(x - j_1) \cdot N_p(y - j_2)$, $X = (x, y)$, where $j_1, j_2$ represent the index coordinates in each dimension corresponding to the $j^{th}$ basis function. Moreover, $N_p(x)$ is the 1D reference B-spline of degree $p$ centered at $x = 0$. This computation is efficient as it does not requires the solving of a linear system of equations. A fast way to compute the coefficients is to employ an image filtering technique. Considering that $g(X)$ is a piecewise constant function, and $N_{j,p}(X)$ is a piecewise polynomial, the numerator of (2) can thus be rewritten as:

$$\begin{aligned}
\int_{\Omega_I} N_{j,p}(X) g(X)\, dX &= \int_{\Omega_I} N_p(x - j_1) N_p(y - j_2) g(X)\, dX \\
&= \sum_{i_1=0}^{p} \sum_{i_2=0}^{p} a_{i_1} a_{i_2} I_{i_1+j_1, i_2+j_2} \\
&= h_a * M_j.
\end{aligned} \tag{3}$$

where $a_i = \int_{i-\lfloor \frac{p+1}{2} \rfloor}^{i-\lceil \frac{p+1}{2} \rceil + 1} N_p(x)\, dx$, and $I_{i_1+j_1, i_2+j_2}$ is the intensity of the voxel corresponding to the multi-index $(i_1 + j_1, i_2 + j_2)$. Moreover, $h_a$ is a convolution matrix given by:

$$h_a(i_1, i_2) = a_{i_1} \cdot a_{i_2}, \quad i_1, i_2 = 0, 1, 2, ..., p, \quad M_j(i_1, i_2) = I_{j_1+i_1, j_2+i_2}. \tag{4}$$

Finally, (2) can be computed efficiently as follows:

$$c_j = \frac{h_a * M_j}{\int_{\Omega_I} N_{j,p}(X)\, dX}. \tag{5}$$

Because $N_{j,p}(X)$ is translation invariant, the denominator of (5) also needs to be computed just once in the interior of the domain. We refer to [31] for a detailed discussion of the properties of this B-spline representation of the domain.

**Remark 1.** This procedure can be trivially parallelized and the convolution is separable which means that it can be first performed in the $x$-direction for each row of voxel, then in the $y$-direction and similarly for higher dimensions. It is also very efficient to evaluate the initial values and the gradient at equally spaced points in the domain by simply pre-computing the value of the B-spline basis at the desired locations on a reference knot span.

We mention that the B-spline kernel used in the convolution is related to the Gaussian filter in image processing applications. As a result of this filter, a noise resistant boundary which is suitable for further computational processing can be obtained. To demonstrate the advantage of obtaining the boundary using B-spline level set functions, we show a comparison of the boundary obtained this way with the boundary obtained directly form the discrete gray scale function. In Figure 1a, we reproduce a spanner image from [30]. The contour obtained by the B-spline level set representation is given in Figure 1c, and Figure 1b is the contour obtained from the gray scale function for a particular intensity threshold. It is clear that the contour in Figure 1c is smoother and less prone to compression artifacts.

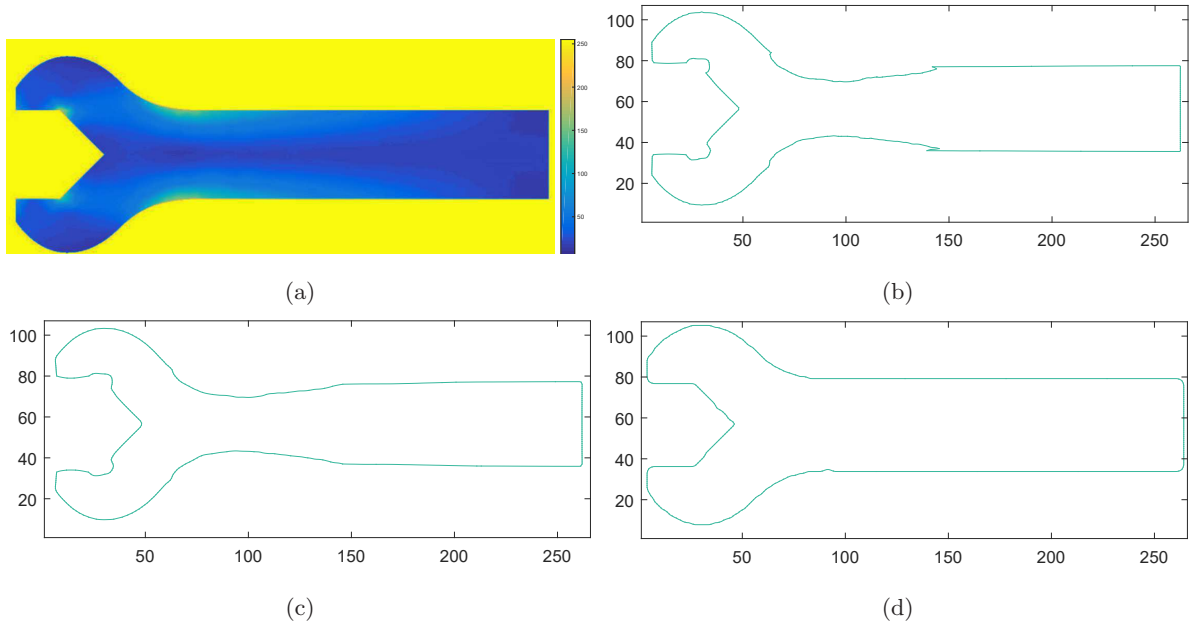

Figure 1: (a) Spanner image, (b) contour from gray scale function, (c) contour from level set function, and (d) contour from level set after resetting the intensity to 0 and 255.

We note that by adjusting the threshold value, a good representation of the domain of interest can be obtained. To further ensure that the gradients around the boundary are predictable and not affected by image noise, a pre-processing step before convolution can be used in the case where the image is not black and white. We set the intensity for all the voxels outside the computational domain to 255, while intensity of the voxels inside the computational domain is set to 0. This essentially converts the gray scale image to a black and white image where black represents the domain of interest. The final boundary representation for this example is shown in Figure 1d.

## 3. Polynomial Splines over Hierarchical T-meshes (PHT Splines)

The polynomial splines introduced in [7] are related to B-splines of reduced continuity over hierarchical T-meshes. Local refinement is an appealing property of the PHT-splines, which makes their use suitable for fitting arbitrary domains.

*3.1. Construction of PHT-splines*

In 1D, the initial PHT-spline representation for a given curve is of the form:

$$F(u) = \sum_{i=1}^{n} B_i(u) \, P_i \,, \tag{6}$$

where $P_i$ are the control points and $B_i(u)$ are the cubic B-spline basis functions defined over the knot vector

$$\Xi^u = \left\{ 0, 0, 0, 0, \frac{1}{n}, \frac{1}{n}, ..., \frac{n-1}{n}, \frac{n-1}{n}, 1, 1, 1, 1 \right\}.$$

The interior knots are repeated once, therefore PHT splines have $C^1$ continuity. For each interior vertex ($u_i$), there are two basis functions supporting $[u_{i-1}, u_{i+1}]$; we call these points *basis vertices*. The associated basis functions are determined by the local knot vectors $(u_{i-1}, u_{i-1}, u_i, u_i, u_{i+1})$, and $(u_{i-1}, u_i, u_i, u_{i+1}, u_{i+1})$ respectively. When a new basis vertex is inserted, there are 2 additional new basis functions, and therefore two new control points need to be calculated, while the previous control points stay fixed. In particular, the refined PHT-spline can be expressed as follows:

$$\hat{F}(u) = \sum_{i=1}^{n} \hat{B}_i^{\ell}(u) \cdot P_i + \sum_{k=1}^{2} \hat{B}_k^{\ell+1}(u) \cdot P_k \,, \tag{7}$$

where $\hat{F}(u)$ is the new geometry parametrization, $\hat{B}_i^{\ell}(u)$ are the (possibly modified) basis functions from the previous level ($\ell$) and $\hat{B}_k^{\ell+1}(u)$ are the new basis functions. If the geometry parametrization is fixed, then $\hat{F}(u) = F(u)$.

In order to modify the basis functions, the Bézier ordinates representation is used. Suppose an element $\phi$ is refined at level $\ell$. The corresponding Bézier ordinates of $\phi$ are subdivided using the DeCasteljau algorithm (see formula (14.15) in [10]) into two parts. The basis function $B_i^{\ell}(u)$ is modified to $\hat{B}_i^{\ell}(u)$ by resetting the Bézier ordinates associated with the new knot to zero. Let the new knot be given by $u_{\hat{i}}$, then the two new basis functions are defined through the knot vector $(u_{\hat{i}-1}, u_{\hat{i}-1}, u_{\hat{i}}, u_{\hat{i}}, u_{\hat{i}+1})$, and $(u_{\hat{i}-1}, u_{\hat{i}}, u_{\hat{i}}, u_{\hat{i}+1}, u_{\hat{i}+1})$ respectively.

Figure 2 illustrates the change of basis functions when a new knot is inserted, which is different from standard B-spline knot insertion. The basis functions in Figure 2a are defined by the knot vector

$$\Xi = \{0, 0, 0, 0, \frac{1}{4}, \frac{1}{4}, \frac{1}{2}, \frac{1}{2}, \frac{3}{4}, \frac{3}{4}, 1, 1, 1, 1\},$$

and Figure 2b shows the modified basis functions when a basis vertex at $\frac{5}{8}$ is inserted. We observe that in this case the only modifications to the basis functions take place on the element that is being refined. Even in higher dimensions, an element can be refined by changing the Bézier representation of the basis on the current element and its immediate neighbor only. The reduced overlap between the basis functions allows for more efficient and localized refinement schemes compared to other local spline bases, such as hierarchical B-splines and T-splines.

We note that in Figure 2, all functions except the modified basis functions which have support on the boundary of the refined knotspan are B-splines. A PHT-mesh can be converted to a B-spline mesh by repeatedly inserting vertices until all the knotspans are at the same refinement level. Also similarly to T-splines, a PHT-spline mesh is determined completely in terms of the local knot vectors and the Bézier ordinates of the basis functions on each knotspan. The definition of basis functions for higher dimension PHT-splines can be generalized using a similar method. For dimension $d$, a new knot (basis vertex) introduces $2^d$ new basis functions.

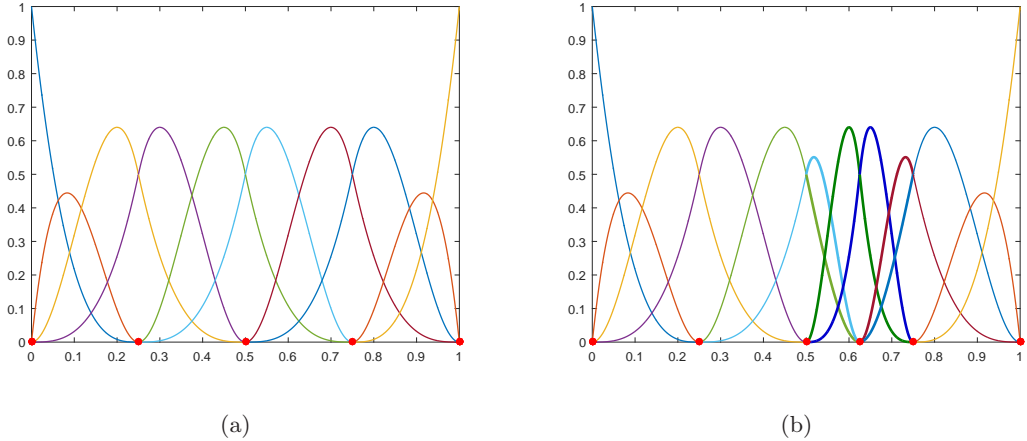

Figure 2: (a) Basis functions generated over $\Xi$, (b) modified and new basis functions when a basis vertex at $\frac{5}{8}$ is inserted.

3.2. PHT-Spline Space

For simplicity, we briefly describe the concept of T-meshes in 2D. Referring to [6], the spline space over a given T-mesh $\mathcal{T}(\mathcal{F})$ is defined as:

$$\mathcal{T}(\mathcal{F}):\, = \{s \in C^{1,1}(\Omega) : s|_\phi \in \mathbb{P}_3 \text{ for any } \phi \in \mathcal{F}\}, \tag{8}$$

where $\mathbb{P}_3$ is the space of all cubic polynomials, $\mathcal{F}$ denotes the set of elements in the mesh, and $C^{1,1}(\Omega)$ is the space consisting of all the functions which have $C^1$ continuity in each spacial direction of $\Omega$.

Refinement is always done by "cross-insertion", i.e. splitting a square element into four elements in 2D and a cube element into eight sub-cubes in 3D. The basis vertices in 2D are those located at the intersection of four elements in the interior, and all the boundaries and corner vertices. The basis vertices in 3D are located at the intersection of eight elements in the interior, four elements on the faces as well as all the edges and corner vertices.

Since the PHT-spline is a cubic spline with $C^1$ continuity, the dimension of the PHT-spline space over the mesh $\mathcal{F}$ is given by:

$$\dim(\mathcal{T}(\mathcal{F})) = 2^d \cdot (\text{number of basis vertices}).$$

We denote the spline space at level $\ell$ as $\mathcal{T}(\mathcal{F}_{\ell+1})$. At the new level, the new spline space $\mathcal{T}(\mathcal{F}_{\ell+1})$ will be changed such that the basis functions inherited from the level $\ell$ will have minimal changes when compared to the basis functions at level $\ell + 1$. In particular, they have the same shape as those in level $\ell$ outside the elements that contain new basis vertices. This can be seen by comparing the shape of basis functions before and after a new vertex is inserted (see Figure 3a and Figure 3b). This is an important feature of the PHT-splines, which reduces the overhead associated with local refinement. The T-vertices (located at T-junctions) have no basis function associated with them. We note that the neighbor of a refined element only needs to be changed when a T-vertex is converted to a basis vertex. This is different from T-splines where the addition of an edge to the mesh can trigger additional refinements further away from the inserted edge. In 3D, the modification of existing basis functions and the computation of new ones can be done using a similar method, taking advantage of the local tensor product structure of the refined element. Compared to 2D, there are eight basis functions supporting each vertex instead of four. Thus, adding a new vertex introduces eight new basis functions to the space. The modification of basis functions and computation of new basis functions can be done similarly.

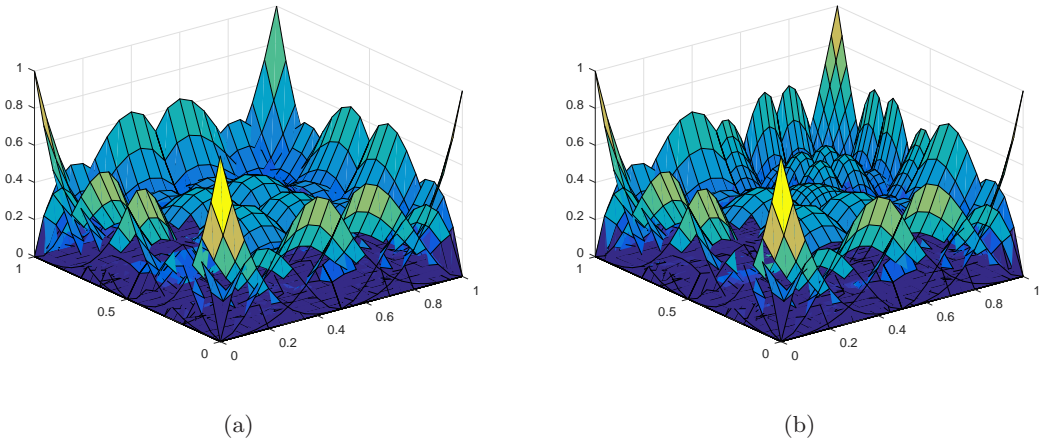

Figure 3: Modification of basis functions in 2D, (a) basis functions at level $\ell$, and (b) basis functions at level $\ell + 1$.

### 3.3. Computing the control points

For a linear mapping between the parameter space and the physical space, the control points of the initial mesh ($\ell = 1$) are set as the locations of the Greville Abscissae, while the control points at level $\ell > 1$ can be computed using the *geometric information* of the basis functions [7]. The geometric information is determined by the location of the basis vertex in the physical space and the values of the derivatives of the mapping evaluated at the basis vertex. In 1D, the geometric information is given by the linear operator

$$\mathcal{L}F(u) = \left(F(u), \frac{\partial F(u)}{\partial u}\right) = (F(u), F_u(u)).$$

Suppose the spline is defined as:

$$F(u) = \sum_{i=1}^{n} B_i(u) \cdot P_i, \tag{9}$$

then the geometric information of the spline at an arbitrary knot $u_a$ denoted ($\mathcal{L}F(u_a)$) is given by:

$$\mathcal{L}F(u_a) = \sum_{j=1}^{2} \mathcal{L}(B_{i_j}(u_a) \cdot P_i)$$
$$= \boldsymbol{P} \cdot \boldsymbol{B}, \tag{10}$$

where $i_1$ and $i_2$ represent the two basis indices corresponding to the knot $u_a$. Note that $\boldsymbol{P}$ and $\boldsymbol{B}$ are 1×2 and 2×2 matrices respectively. Hence, $\boldsymbol{P}$ can be solved from:

$$\boldsymbol{P} = \mathcal{L}F(u_a) \cdot \boldsymbol{B}^{-1}. \tag{11}$$

This method can be extended to finding $\boldsymbol{P}$ when dealing with higher dimensions. We refer to [7] for an explanation of computing $\boldsymbol{P}$ in 2D. Here we discuss the computation of control points in 3D. The geometric information in 3D is defined as follows:

$$\mathcal{L}F(u,v,w) = [F(u,v,w), \frac{\partial F(u,v,w)}{\partial u}, \frac{\partial F(u,v,w)}{\partial v}, \frac{\partial F(u,v,w)}{\partial w},$$
$$\frac{\partial F^2(u,v,w)}{\partial u \partial v}, \frac{\partial F^2(u,v,w)}{\partial u \partial w}, \frac{\partial F^2(u,v,w)}{\partial v \partial w}, \frac{\partial F^3(u,v,w)}{\partial u \partial v \partial w}]. \tag{12}$$

Thus, for a basis vertex $(u_a, v_a, w_a)$, the control points can be obtained from the relation:

$$\boldsymbol{P} = \mathcal{L}F(u_a, v_a, w_a) \cdot \boldsymbol{B}^{-1}. \tag{13}$$

Taking advantages of the closed form of B-splines and assuming the new basis vertex is in the middle of the refined element, elementary operations show that $\boldsymbol{B}$ is a 8×8 matrix given by:

$$\boldsymbol{B} = \boldsymbol{W} \otimes [\boldsymbol{V} \otimes \boldsymbol{U}], \tag{14}$$

where $\otimes$ is the Kronecker product. Moreover $\boldsymbol{U} = \begin{bmatrix} (1-\lambda) & \lambda \\ -\alpha & \alpha \end{bmatrix}$, $\boldsymbol{V} = \begin{bmatrix} (1-\mu) & \mu \\ -\beta & \beta \end{bmatrix}$, $\boldsymbol{W} = \begin{bmatrix} (1-\nu) & \nu \\ -\gamma & \gamma \end{bmatrix}$, where $\alpha = \frac{1}{\Delta u_1 + \Delta u_2}$, $\beta = \frac{1}{\Delta v_1 + \Delta v_2}$, $\gamma = \frac{1}{\Delta w_1 + \Delta w_2}$, $\lambda = \alpha \Delta u_1$, $\mu = \beta \Delta v_1$, and $\nu = \gamma \Delta w_1$. $\Delta u_i$, $\Delta v_i$ and $\Delta w_i$, $i = 1, 2$ are the differences between $u_a, v_a,$ and $w_a$ and the neighbor knots in the negative and positive direction respectively.

We will describe boundary approximation of an image using PHT-spline in the following section. In brief, we preform iterative refinements and adjust the boundary vertices to better match the geometry of the domain. The control points corresponding to the new basis vertices are first calculated as described in this subsection.

## 4. Volumetric parametrization of image data

We propose a method to obtain a smooth boundary representation of an image using PHT-splines by refining the boundary elements and also by adjusting the associated vertices. For a given image, we first obtain the B-spline level set representation $f(X)$ of its boundary. We denote the physical domain by $\Omega_{phys}$, and embed it in a domain $\Omega$ which has a much simpler geometry. In 2D, we consider $\Omega$ to be a rectangular tensor-product mesh, while in 3D, a cuboid is used as $\Omega$. Next, we obtain the level set representation of the image and set the points where the level set value is below the threshold $T = 200$ as the physical domain $\Omega_{phys}$. This is followed by embedding $\Omega_{phys}$ in $\Omega$ with a resolution which at least captures the coarse details of the image, and proceed with the Algorithm 1. Each step in Algorithm 1 is presented in more detail in the following subsections.

---

**Algorithm 1:** Volumetric parametrization

**Input** : PHT-mesh, $\Omega$ embedded with $\Omega_{phys}$, level set function ($f(X)$), intensity threshold $T$, P%

**Output:** Modified PHT-mesh, where the boundary of active domain $\Omega_{active}$ matches the boundary of $\Omega_{phys}$

1: Determine and classify boundary elements. (Algorithm 2 in Appendix)
2: Determine adjustable vertices. (Algorithm 3 in Appendix)
3: Adjust vertices using the circle or sphere template. (Algorithm 4 in Appendix)
4: Smooth and untangle distorted elements.
5: If a better boundary representation is desirable, refine the boundary elements and their neighbors, and repeat from 1.

---

*4.1. Classifying the boundary elements*

We classify the elements in $\Omega$ into 4 groups as follows:

- Boundary element of type 1 are elements cut by the contour of $\Omega_{phys}$ and more than P% of the volume has intensity less than $T$.

- Boundary element of type 2 are elements cut by the contour of $\Omega_{phys}$ and less than P% of the volume has intensity greater than or equal to $T$.

- Internal elements are elements inside the physical domain.

- External elements are elements outside the physical domain.

The parameter P% is chosen based on the image data in order to prevent very small elements from being considered in the computational domain. A value of $P = 50$ works well in most cases. The classification of each element is illustrated in Figure 4a, with the red color cells, green color cells, blue color cells and white color cells representing Boundary elements of type 1, Boundary elements of type 2, internal elements and external elements respectively.

The goal is to obtain an active domain $\Omega_{active} \subset \Omega$ such that $\Omega_{active} \approx \Omega_{phys}$. The $\Omega_{active}$ consists of boundary elements type 1 and internal elements, i.e. $\Omega_{active} = \Omega_{B1} \cup \Omega_{In}$. Our proposed approach is to deform the boundary elements of type 1 and 2 such that the boundary of $\Omega_{active}$ matches the boundary of $\Omega_{phys}$. We note that the exterior elements and elements of type 2 are classified as *inactive elements*. While they are part of the hierarchical mesh and have associated control points, these elements are not considered during the assembly and solution of the given PDE.

*4.2. Determining the adjustable vertices*

After classifying the elements, we select the *adjustable vertices* from the boundary elements as follows:

- A vertex belonging to a Boundary element of type 1 is marked "adjustable" if the intensity value at the corresponding location is greater than or equal to the threshold $T$ (it lies outside $\Omega_{phys}$).

- A vertex belonging to a Boundary element of type 2 is marked "adjustable" if the intensity value at the corresponding location is less than the threshold $T$ (it lies inside $\Omega_{phys}$).

We would like to adjust (move) the selected vertices so that they touch the boundary of $\Omega_{phys}$, hence enforcing that the boundary of $\Omega_{active}$ resembles the boundary of physical domain. The control points associated to the vertices are adjusted by solving either a linear system or by using a predefined template for the unit circle or sphere, as shown in the following subsections.

*4.3. Adjusting vertices*

In our proposed method, the vertices are moved to the boundary in a similar way as presented in [27] and [36]. In [27], the mesh is modified by weighting the B-splines so that mesh vertices of the boundary elements move to the closest location on the boundary. In [36], the vertices are first smoothed before they are projected in the normal direction onto the boundary using the Newton-Raphson method. The control points are then recovered by an interpolation method which involves solving a linear system of equations.

The proposed method intends to move the vertices of the boundary elements by taking advantage of the fact that each basis vertex is associated with $2^d$ control points. We move the vertices by assigning new locations to the corresponding control points, which determine locally the geometric parametrization. The locations of the control points and mesh geometry are obtained from the translated, scaled, and rotated circle or sphere template. Therefore, an advantage of the proposed method is that the geometric information of the level set function (such as normal and curvature) can be approximated by the PHT-splines.

*4.3.1. Determining the new vertex location*

We move an adjustable vertex along the edge or diagonal of an element according to type of elements that they are connected to. In 2D, if the adjustable vertex is shared by two boundary elements of the same type, we adjust it along the shared edge of these elements otherwise it is adjusted along the diagonal direction. A 2D example of the proposed method is illustrated in Figure 4. In the first figure (Figure 4a), we embed the level set representation of the image into an uniform mesh and classify the boundary elements. The Boundary element type 1, Boundary element type 2, and internal elements are highlighted with red, green and blue color respectively. The union of the red and the blue elements forms the active domain, which will be used for analysis. Adjustable vertices are marked with black dots and squares, where the dots are the adjustable vertices of Boundary element type 1, and squares are the adjustable vertices of Boundary element type 2. Figure 4b shows the result after adjustment. In 3D, the determination of the adjusted direction is

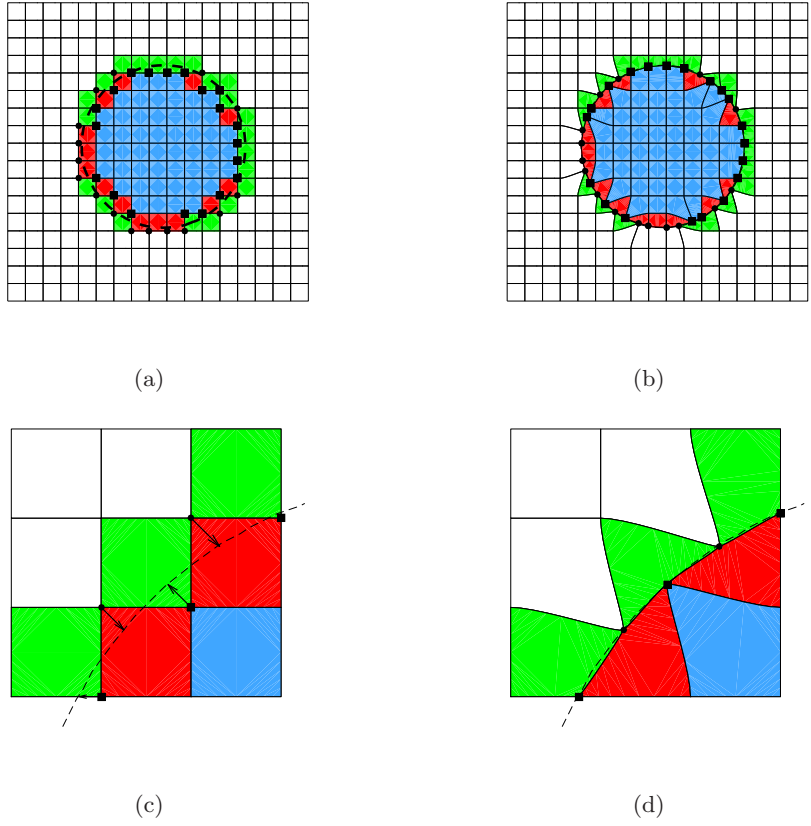

Figure 4: 2D example of the boundary matching algorithm, (a)initial mesh and element classification, (b) after adjusting the selected vertices of boundary elements, (c) zoomed in view of the adjusted vertices and (d) the resulting bounding after adjustment.

classified into 3 cases as follows:

1. Adjustable vertex belongs to only one boundary element (see Figure 5a).

2. Adjustable vertex shared by two adjacent boundary elements (see Figure 5b).

3. Adjustable vertex shared by four boundary elements (see Figure 5c).

In Figure 5, the black lines represent the element edges, the blue dots are the level set contour (boundary of the physical domain) cutting through the elements, and the black dots are the adjustable vertices. Figure 5a illustrates the case when the adjustable vertex belongs to only one element. In this case, we adjust the vertex along the cube diagonal. In Figure 5b, the adjustable vertex is shared by two adjacent elements, and the shared face diagonal is selected as the adjustable direction. For the third case, where the adjustable vertex is shared by four elements and these four elements share an edge, as shown in Figure 5c, we move the adjustable vertex along the shared edge. In all these figures, the red arrow represents the moving direction of the adjustable vertex, and the black dashes represent the diagonal. After determining the adjustable direction, we apply the bisection method to compute the cut location of the physical boundary at the selected edge or diagonal. In 3D, for the adjustable vertices which do not belong to any of these three cases, we set the nearest point on the physical boundary as the cut location.

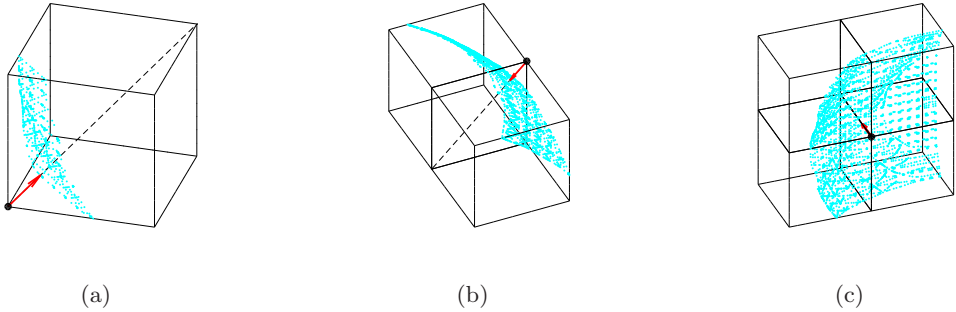

(a) (b) (c)

Figure 5: 3D classification of adjust direction (a) vertex belongs to one boundary element, (b) vertex shared by two adjacent elements, and (c) vertex shared by four elements.

*4.3.2. Circle and sphere templates*

In 2D, suppose $(u_a, v_a)$ is an adjustable vertex; there are four basis functions associated with it. The geometric information of this vertex can be written in the matrix form similarly to (10). We would like to find the coefficients ($\boldsymbol{P}$) such that at the vertex, the outer normal of the boundary edge matches that of the contour line of the level set. The normal of an element in the parameter space and the normals in the physical space are related to the first derivatives of the mapping $F$ by Nansen's formula, which is often used when imposing Neumann boundary conditions in isogeometric analysis. We note that there are several cases depending on whether the adjustable vertex is a corner vertex of the active domain or if it is shared by more than one active elements. Also, the magnitude of the normal vectors and the values of the cross derivatives of the mapping (i.e. $\frac{\partial F}{\partial u \partial v}$), which correspond to the "corner twists" of the element, are free parameters which need to be estimated. In [7], a local parametrization using quadratic polynomials is suggested as the basis for extracting the geometric information. This procedure may however be computationally expensive. For these reasons, we propose to use a template that is scaled and translated to each element, as shown in Figure 6.

A good way to analyze the shape of a curve is to use the osculating circle. The osculating circle at a point $x$ on a curve is the circle that approximates the curve by matching its tangent and curvature [11]. Instead of solving a linear system, we also propose to use the corresponding control points from a circle template. The resulting control points will generate an arc passing through the adjusted vertices and approximating the geometric information of the physical boundary. First, we generate a circle template that approximates the unit circle. This template is formed by 16 elements, and the approximating circle is composed by four elements labeled 6, 7, 10 and 11 (see Figure 6a). In the template, the control points and the vertices are represented by the red dots and black dots respectively, and the circular arcs are approximated by the edges highlighted with green color. We note that some of the control points are repeated, which results in a singularity in the geometric mapping and may cause problems in analysis. We can avoid this problem by slightly separating the repeated control points. The control points of the element labeled 10 before and after the separation are shown in Figure 6c and 6d respectively. In these figures, the black dots represented the vertices, and the associative control points of each vertex are marked with different colors as: Vertex 1 - green , Vertex 2 - purple, Vertex 3 - red, and Vertex 4 -cyan. In 3D, we approximate the boundary using a sphere template as shown in Figure 7. The distance error from the actual unit sphere is plotted, and we can see that even with eight elements the error is relatively low.

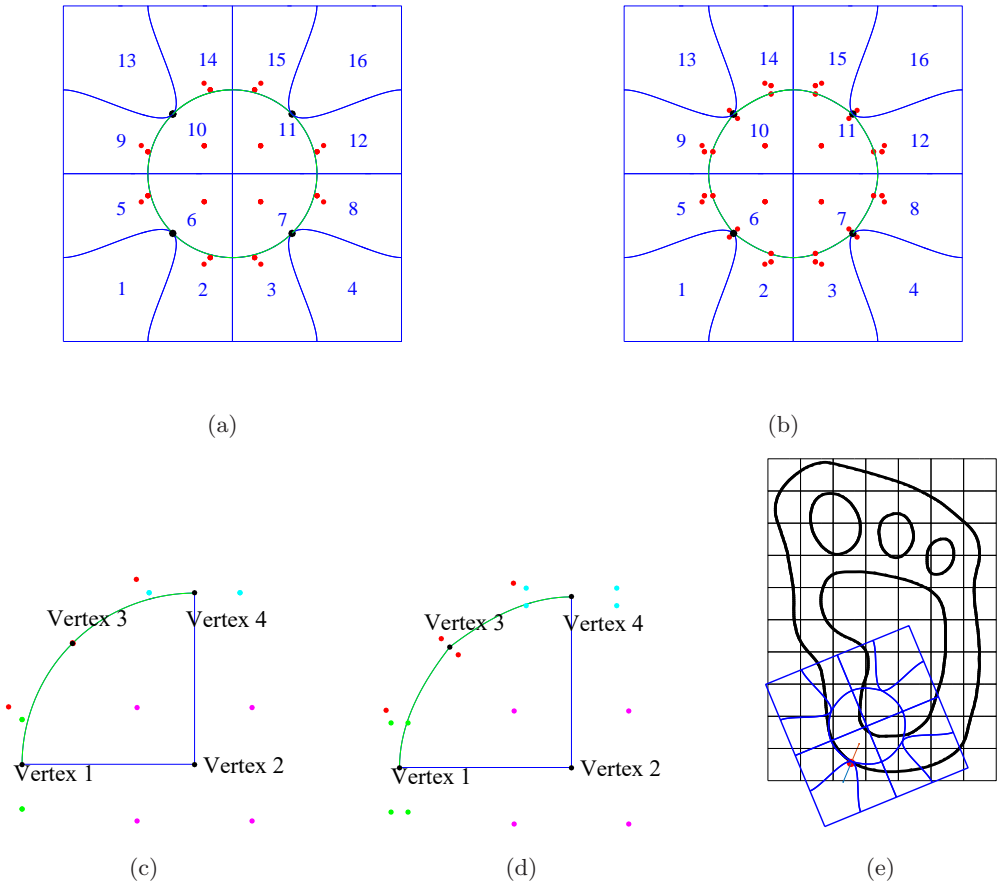

Figure 6: Osculating circle example, (a) circle template with repeated control points, (b) alternate circle template, (c) element 10 with repeated control points, (d) element 10 with separated control points, and (d) Yeti footprint and a scaled and rotated circle template at a selected contour point.

To use the circle template, we start by computing the curvature ($\kappa$) and normal direction at the cut location from the level set function. Next, we generate a circle using the template by setting the radius $r = 1/\kappa$. The circle is then rotated such that the normal at the corresponding vertex matches with the normal obtained from the level set. A good boundary approximation can be obtained by replacing the associated control points with those obtained from the rotated and scaled circle template. Figure 6e illustrates how a selected part of the contour is approximated by using the circle template. The circle template is scaled and rotated to match with the geometric information at the given point (represented by a dot in the 6e) of the contour, and we will approximate the boundary with an arc of the circle template. We note that the scaling should take into account both the curvature at the boundary and the size of the element to avoid fold-overs. We use the sphere template in a similar way to the circle template.

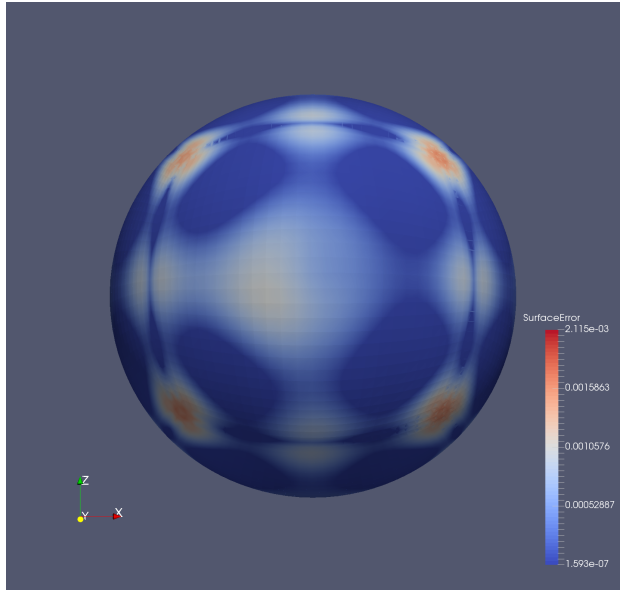

Figure 7: Sphere template (the interior 8 elements which approximate the unit sphere are shown) and surface error.

We mention that the adjustment may lead to distorted elements and non-regular points. However, the mesh quality can be improved with a smoothing operation which will be discussed in Section 4.4. In addition, a better approximation to the boundary can be obtained by refining the boundary elements. The refinement strategy is discussed in the Section 4.5.

*4.4. Smoothing and untangling the elements*

In some cases, moving the vertices to the cut locations may lead to distortion. To address this problem, we apply a Laplacian smoothing operation after adjusting the vertices. Suppose $P_j$ and $P_i$ are the positions of $j$ and $i$ vertices respectively, while $\hat{P}_i$ denotes the new position of vertex $i$. The smoothing operation is described as:

$$\hat{P}_i \leftarrow P_i + \lambda L(P_i), \; 0 < \lambda < 1 \tag{15}$$

where $\lambda$ is a regularization term, and $L(P_i)$ is the Laplacian smoothing defined as

$$L(P_i) = \frac{1}{N_i} \sum_{j=1}^{N_i} P_j - P_i. \tag{16}$$

where $N_i$ represents the number of neighbors to the vertex $i$. In the numerical experiments shown below, we found that $\lambda = 0.1$ works well in most cases.

*4.5. Refinement*

In order to obtain a better boundary representation, a multi-level refinement is performed along the boundary elements. For a given PHT-mesh, which is generated according to Section 3, refinement at each level can be generated by successively inserting new basis vertices onto the parametric domain of the boundary elements. Referring to Section 4.3, the vertices are adjusted by translating, scaling and rotating the associated control points. Since there are no specific control points associated to a vertex at the T-junction, it is preferable to avoid T-junctions on the boundary. Thus, in addition to refining the boundary elements, the neighbor elements are refined as well. On each refinement level, we carry out the following steps:

1. Determine Boundary elements of type 1 and Boundary elements of type 2.

2. Determine the elements bounding both types of Boundary elements.

3. Insert new basis vertices in the elements obtained from 1 and 2 as discussed in Section 3.1.

## 5. Numerical Examples

We illustrate the accuracy and efficiency of the proposed method with several 2D and 3D examples. The first example is a synthetic benchmark problem - a hole in an infinite plate. The second example is the Yeti footprint example used to demonstrate the efficiency of the proposed method for handling complex shapes. The third example is an open spanner example where we show the application to elastostatic analysis. We examine the proposed method in 3D using the spherical hole in an infinite domain benchmark, and a problem using the Stanford bunny as a computational domain is shown as the last example. For those examples involving analysis, the Young's modulus is set as $10^5$ and the Poisson's ratio ($\nu$) is set as 0.3.

*5.1. Hole in an infinite plate*

The example of the hole in an infinite plate subject to uniaxial tension is shown in Figure 8. The exact solution of this problem is given by:

$$\begin{aligned}
\sigma_{rr} &= \frac{\sigma_\infty}{2}\left(1 - \frac{R^2}{r^2}\right) + \frac{\sigma_\infty}{2}\left(1 - \frac{4R^2}{r^2} + \frac{3R^4}{r^4}\right)\cos 2\theta \\
\sigma_{\theta\theta} &= \frac{\sigma_\infty}{2}\left(1 + \frac{R^2}{r^2}\right) - \frac{\sigma_\infty}{2}\left(1 + \frac{3R^4}{r^4}\right)\cos 2\theta \\
\sigma_{r\theta} &= -\frac{\sigma_\infty}{2}\left(1 + \frac{2R^2}{r^2} - \frac{3R^4}{r^4}\right)\sin 2\theta,
\end{aligned} \tag{17}$$

where $\sigma_\infty$ is the remote stress. $R$ is the hole radius and $r$ and $\theta$ are polar coordinates with respect to the center of the hole. In this example, we set $\sigma_\infty = 10$ and $R = 1$. A PHT-spline representation of the active domain of the plate is shown in Figure 8a, and the analysis result is illustrated in Figure 8b. The analysis result can be improved when more refinement steps are performed, as shown in Figure 8d. Figure 8c shows a comparison of the convergence rates. The red lines shows the convergence of the error in the energy norm for this problem when we refine the mesh and also adjust the vertices along the boundary using the proposed method. The blue line shows the result when only uniform refinement is carried out. We can see that after a few refinements, the convergence rate levels off. The convergence rate when the vertices are moved to the curved boundary at each refinement step is stable and close to the reference rate which is represented by the dashed line.

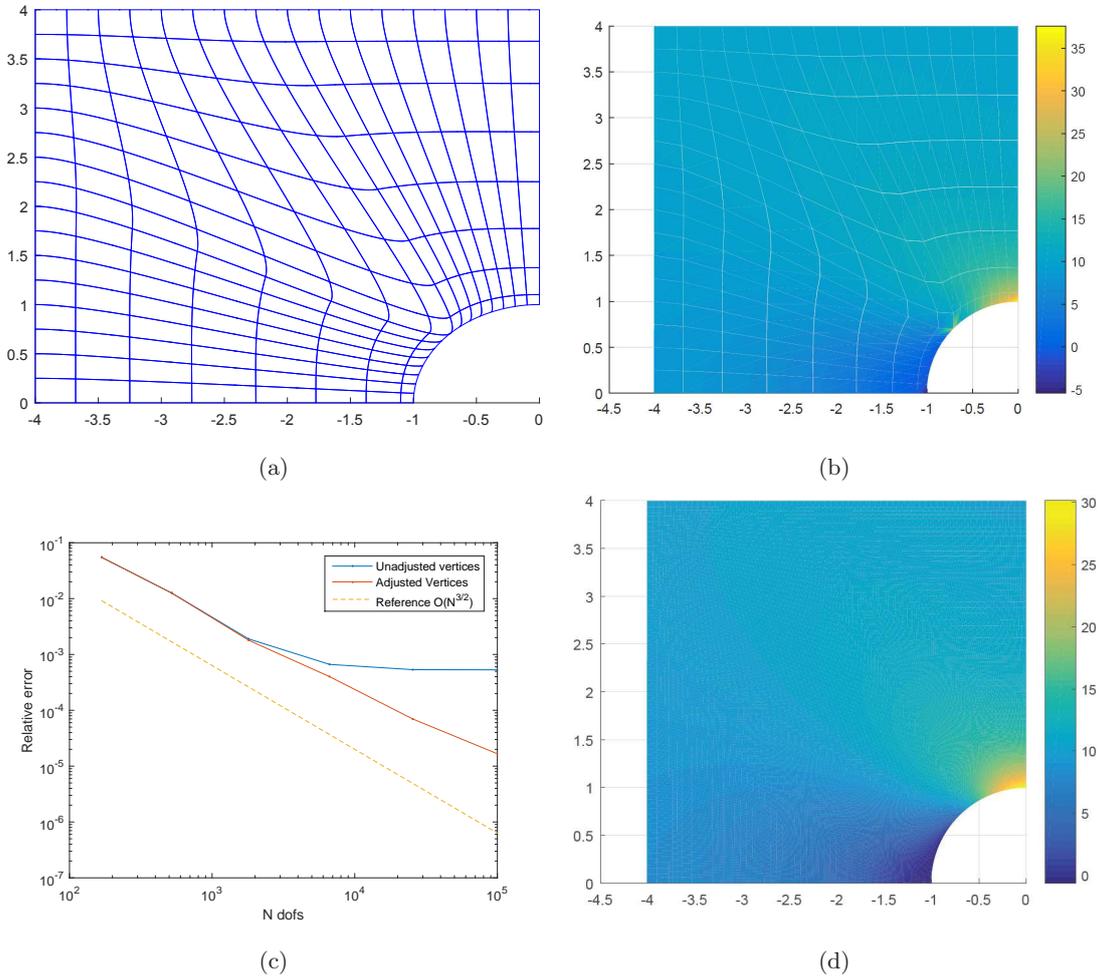

Figure 8: Synthetic benchmark: plate with a hole (a) coarse mesh, (b) $\sigma_{xx}$ component of the stress field, (c) convergence plot for the energy norm comparing the error in approximation for the fixed geometry vs. using vertex adjustment near the curved part of the domain, (d) $\sigma_{xx}$ stress on a finer mesh.

## 5.2. Yeti footprint

To demonstrate the ability of the proposed method in handling complicated shapes, including those with holes or cracks, we use the Yeti footprint as the next example. In Figure 9, we illustrate the representation of Yeti footprint with only one PHT-spline patch. Comparing to the example shown in [17], where 21 patches are used to form the domain, the proposed method is more straight-forward. Figure 9a shows the Yeti footprint image obtained from [17]. Figure 9b is the result of approximating the Yeti image with a coarse mesh, where we can see that the boundary is already well-resolved. However, if a better approximation is needed, the boundary representation can be further enhanced with additional refinements. The result of a finer mesh is shown in Figure 9c. The scaled Jacobians of the last mesh are plotted in Figure 9d.

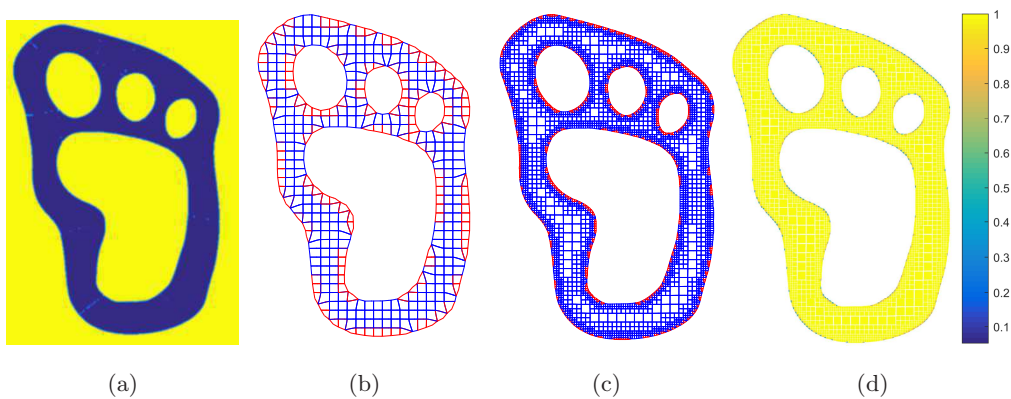

Figure 9: Yeti footprint example, (a) Yeti footprint image, (b) active area of the PHT-mesh after adjusting vertices, (c) the result using a finer mesh, and (d) the scaled Jacobians of the finer mesh.

## 5.3. Open spanner

In this example, we demonstrate an application of the proposed method to elastostatic analysis. We show the open spanner example in the Figure 10. Figure 10a is the spanner image obtained from [30]. The volumetric representation of the spanner using adaptive refinement is shown in Figure 10b, where we can see that sharp corners can be captured better by refining the elements around the corners. Figure 10c is the result of volumetric representation using B-splines; the spanner can be represented well, but, it contains many elements. Thus, the computation is time consuming. The result of uniform refinement along the boundary is shown in Figure 10d. It is clear that by refining only the boundary we can obtain a good boundary approximation, and since there are fewer elements and fewer degree of freedom, the computation is more efficient. The zoomed-in figures of the upper and lower corners at the end of the spanner are shown in Figures 10e and 10f respectively. The analysis results when the corners of the jaw are fixed and uniform traction is applied at the right edge of the handle is shown in Figure 10g, while Figure 10h shows the scaled Jacobians of the parametric mesh.

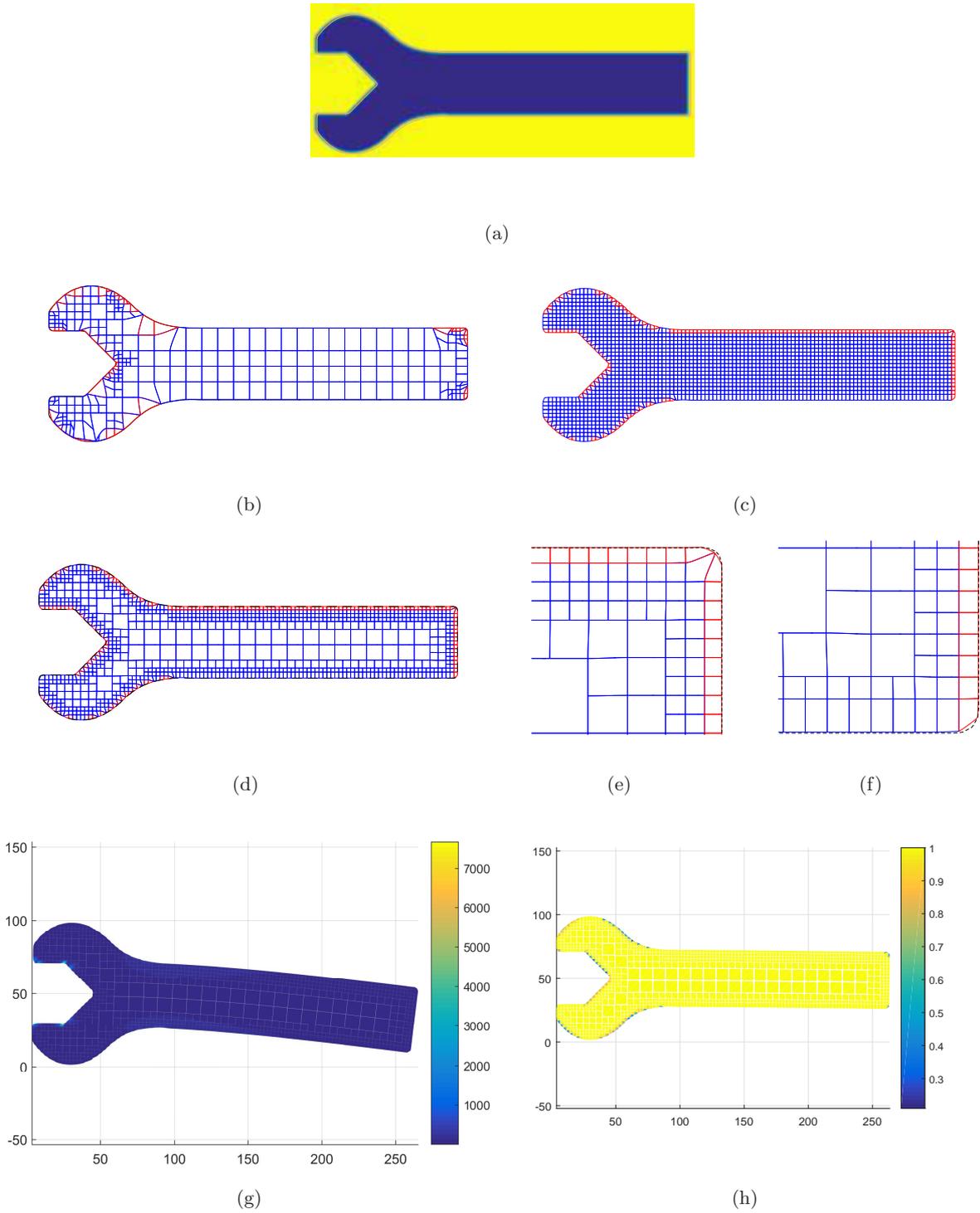

Figure 10: Open spanner example, (a) spanner image, (b) result of adaptive refinement, (c) result of volumetric representation using B-splines, (d) results of uniform refinement along boundary, (e) zoomed-in figure of the upper corned at the end of the spanner, (f) zoomed-in figure of the lower corner at the end of the spanner (g) the analysis result, and (h) the scaled Jacobians.

*5.4. Spherical hole in an infinite domain*

The problem of a spherical hole in an infinite domain is shown is this section. In this problem, an uniaxial tension is applied at infinity in the domain, and the exact stresses are given in terms of spherical coordinates $(R, \theta, \beta)$ as follows [3]:

$$\begin{aligned}
\sigma_{RR} &= \sigma_\infty \cos^2 \beta + \frac{\sigma_\infty}{(7-5\nu)} \left( \frac{a^3}{R^3}(6 - 5(5-\nu)\cos^2\beta) + \frac{6a^2}{R^5}(3\cos^2\beta - 1) \right) \\
\sigma_{\theta\theta} &= \frac{3\sigma_\infty}{2(7-5\nu)} \left( \frac{a^3}{R^3}(5\nu - 2 + 5(1-2\nu)\cos^2\beta) + \frac{a^5}{R^5}(1 - 5\cos^2\beta) \right) \\
\sigma_{\beta\beta} &= \sigma_\infty \sin^2\beta + \frac{\sigma_\infty}{2(7-5\nu)} \left( \frac{a^3}{R^3}(4 - 5\nu + 5(1-2\nu)\cos^2\beta) + \frac{3a^5}{R^5}(3 - 7\cos^2\beta) \right) \\
\sigma_{R\beta} &= \sigma_\infty \left( -1 + \frac{1}{7-5\nu} \left( -\frac{5a^3(1+\nu)}{R^3} + \frac{12a^5}{R^5} \right) \right) \sin\beta \cos\beta
\end{aligned} \quad (18)$$

where $a$ represents the radius of the spherical hole and $\sigma_\infty$ is the uniaxial tension at infinity. Since it is not feasible to model an infinite domain, we consider the solution on part of the domain as in Figure 11a. We compare our results with results obtained from uniform refinement without changing the geometry. The comparison is shown in Figure 11b. The red line is the result of refinement with changing the geometry by adjusting the vertices at the boundary, while the blue line is the result from refinement without adjusting the vertices. We can see that the result is better when we change the geometry at the boundary to approximate the desired shape.

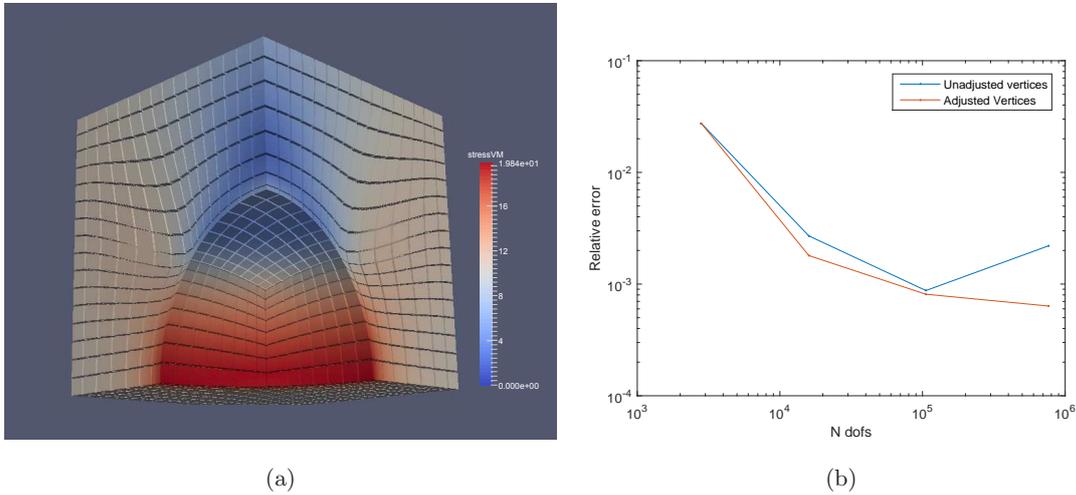

Figure 11: Cube with a hollow sphere example: a) The von Misses stresses on a coarse mesh, b) Convergence plots for the energy norm.

*5.5. Stanford Bunny*

In this example, we demonstrate the ability of the proposed method to handle complex 3D objects. We use the CT-scan data of the Stanford bunny given in [1]. We resize the data to $512 \times 512 \times 512$ voxels such that the aspect ratio of the bunny is scaled to its proper size. Thus, the initial data has approximately 134 million voxels. We consider an approximation of the domain with $64 \times 64 \times 64$ elements. After pre-processing the bunny image to obtain the level set function, the active region of the solid spline is now formed by only 13,754 hexahedral elements. Because the bunny is hollow, most of the domain is formed by inactive elements which do not contribute to the computations. The ability to model objects with voids is one the advantages of our proposed method. In addition, the resulting spline model uses only one patch. Figure 12a and Figure 12b shows the bunny and its base respectively. The volumetric representation and the scaled Jacobians are shown in Figure 12c. The analysis result is illustrated in Figure 12d, where we apply the displacement at one of the ears and fix the base.

## 6. Conclusions

We have presented a self-contained method for converting a voxel-based domain to a spline representation that is suitable for analysis using a hierarchical basis of $C^1$ cubic polynomials. While the continuity of the representation is reduced compared to the highest possible for a cubic basis, the reduced overlap between the basis functions allows for more flexibility in the refinement and boundary representation. The numerical results for benchmark problems show that highly accurate results can be obtained due to the improved geometry approximation. We also show the method is suitable for complex 2D and 3D domains which contain voids or irregular geometric features. Using hierarchical bases that allow for higher continuity in the representation while at the same time preserving sharp features such as corners is planned for future work.

## Acknowledgements

The authors would like the acknowledge the financial support of the German Academic Exchange Program (DAAD) and of the European Commission under the project ERC-COMBAT.

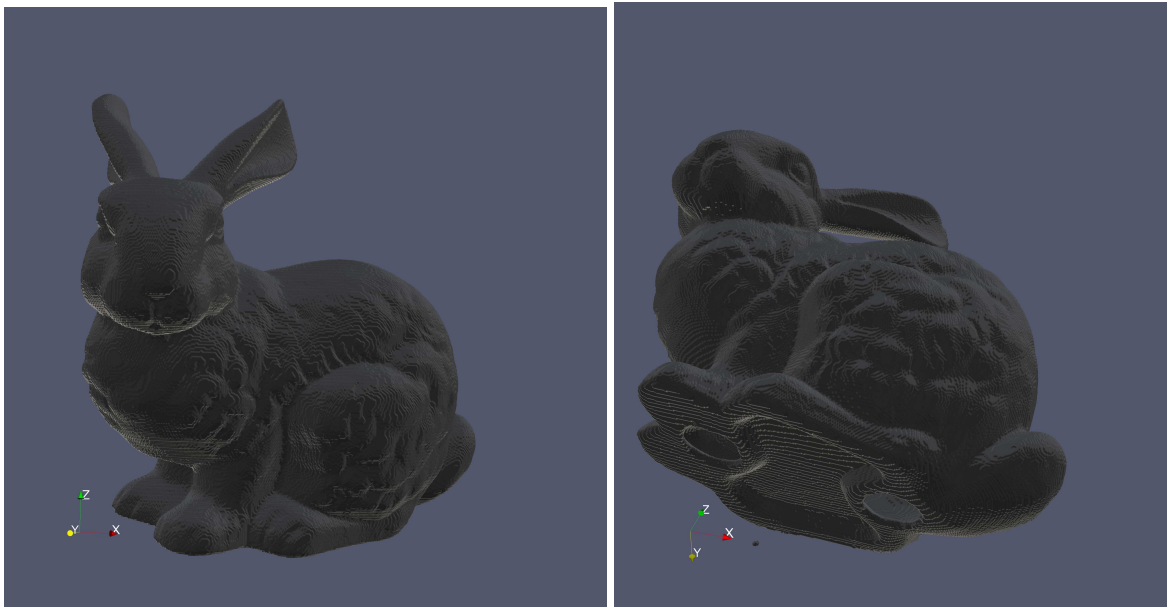

(a)　　　　　　　　　　　　　　　　(b)

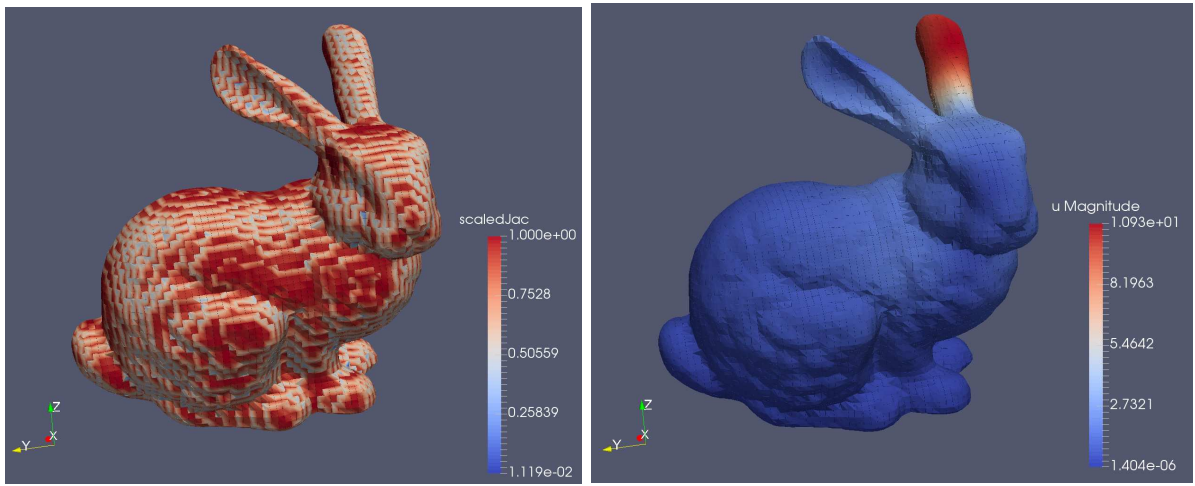

(c)　　　　　　　　　　　　　　　　(d)

Figure 12: Stanford bunny example, (a) level set representation of the Stanford Bunny, (b) the bottom part of the bunny, (c) solid splines representation of the bunny, and (d) the analysis result.

**Appendix**

---
**Algorithm 2:** Determining and classifying boundary element
---
**Input** : PHT-mesh, level set function ($f(X)$), intensity threshold $T$, P%

**Output:** Boundary elements

1: Determine equally-spaced sample points in $\mathcal{F}$.
2: For the sample points, compute the intensity value of these points using $f(X)$.
3: If the intensity value is less than given intensity threshold $T$, set the point as an internal point, otherwise, set it as an external point.
4: Compute the number of internal points and external points within an element.
5: If more than $P$% of the points are internal points, set the element as Boundary element type 1.
6: If less than $P$% of the points are internal points, set the element as Boundary element type 2.
7: If all points are internal points, set the element as Internal element.
8: If all points are external points, set the element as External element.
9: Set the Boundary element type 1 and the Internal elements as the active elements.

---

---
**Algorithm 3:** Determining adjustable vertices
---
**Input** : Boundary element type 1 or 2

**Output:** Adjustable vertices, intensity threshold $T$

1: Compute the intensity value at each vertex.
2: For Boundary element type 1, mark the vertices with intensity value greater than $T$ as adjustable vertices.
3: For Boundary element type 2, mark the vertices with intensity value less then $T$ as adjustable vertices.

---

**Algorithm 4:** Adjusting vertices

**Input** : Adjustable vertex

**Output:** The new locations of the control points associated with the vertex

1: Find the adjustable direction.
2: Compute the cut location using the bisection method, and assign it as the new location of the corresponding vertex.
3: Compute normal and curvature at the new location from the level set function $f(X)$
4: Scale and rotate the circle template according to the normal and curvature values, and translate it to the new location.
5: Relocate the control points associated to the adjustable vertex according to those obtained from the circle or sphere template.